\documentclass[12pt,a4paper]{article}
\usepackage{amssymb,amsmath}
\usepackage{color} 
\usepackage{graphicx}

\newcommand{\bc}{\begin{center}}
\newcommand{\ec}{\end{center}}
\newcommand{\bd}{\begin{displaymath}}
\newcommand{\ed}{\end{displaymath}}
\newcommand{\be}{\begin{equation}}
\newcommand{\ee}{\end{equation}}
\newcommand{\ba}{\begin{array}}
\newcommand{\ea}{\end{array}}
\newcommand{\bt}{\begin{tabular}}
\newcommand{\et}{\end{tabular}}

\begin{document}

\begin{titlepage}


\begin{center}
{
\sffamily
\LARGE
On the suppression of the dark matter-nucleon\\[2mm]
scattering cross section in the SE$_6$SSM}\\[8mm]
{\large Roman Nevzorov\\[3mm]
\itshape{I. E. Tamm Department of Theoretical Physics,}\\[0mm]
\itshape{Lebedev Physical Institute, Leninsky prospect 53, 119991 Moscow, Russia}
}\\[1mm]
\end{center}
\vspace*{0.75cm}

\begin{abstract}{
\noindent
In the $E_6$ inspired $U(1)_N$ extension of the minimal supersymmetric (SUSY) standard model (MSSM)
a single discrete $\tilde{Z}^{H}_2$ symmetry permits suppressing rapid proton decay and non-diagonal
flavour transitions. If matter parity and $\tilde{Z}^{H}_2$ symmetry are preserved this SUSY
model (SE$_6$SSM) may involve two dark matter candidates. In this article we study a new modification
of the SE$_6$SSM in which the cold dark matter is composed of gravitino and the lightest
neutral exotic fermion. We argue that in this case the dark matter-nucleon scattering cross section
can be considerably smaller than the present experimental limit.}
\end{abstract}

\end{titlepage}

\newpage
\section{Introduction}

The possible unification of all interactions remains one of the most appealing
motivations for studying supersymmetric (SUSY) extensions of the Standard Model (SM).
At very high energies, the minimal supersymmetric standard model (MSSM) and its extensions
can be embedded into Grand Unified Theories (GUTs). In $N=2$ SUSY GUTs based on the $E_8$ gauge group
all SM bosons and fermions can belong to a single $248$ representation of $E_8$.
This representation involves three $27 \oplus \overline{27}$, which are fundamental and
antifundamental representations of $E_6$, adjoint $78$ representation of $E_6$ and
$8$ components that do not participate in the $E_6$ gauge interactions. In SUSY GUTs three fundamental
representations of $E_6$ may include Higgs doublet and three families of the SM fermions.
The SM gauge bosons are contained in the adjoint representation of $E_6$.
It is expected that near the scale $M_{0}\gtrsim M_X\sim 2\cdot 10^{16}\,\mbox{GeV}$ the extended SUSY
is broken down to $N=1$ supersymmetry while the breakdown of $E_6$ (or $E_8$) gauge group can lead to
\begin{equation}
\begin{array}{c}
E_6\to SO(10)\times U(1)_{\psi}\to SU(5)\times U(1)_{\chi}\times U(1)_{\psi} \to \qquad\qquad\qquad\qquad\\
\qquad\qquad\qquad\qquad \to SU(3)_C\times SU(2)_W\times U(1)_Y\times U(1)_{\psi}\times U(1)_{\chi} \,,
\end{array}
\label{1}
\end{equation}
where $SU(3)_C\times SU(2)_W\times U(1)_Y$ is a SM gauge group.

The exceptional supersymmetric standard model (E$_6$SSM) \cite{King:2005jy,King:2005my} (for recent review,
see Ref.~\cite{e6ssm-sym}) implies that around the GUT scale $M_X$ rank--6 model with two extra symmetries $U(1)_{\psi}$
and $U(1)_{\chi}$ is reduced further to an effective rank--5 model based on the SM gauge group
together with an additional $U(1)_{N}$ factor, i.e.
\begin{equation}
U(1)_{\psi}\times U(1)_{\chi}\to U(1)_{N}\times Z_{2}^{M}\,,
\label{2}
\end{equation}
where $Z_{2}^{M}=(-1)^{3(B-L)}$ is the so-called matter parity, which is a discrete subgroup
of $U(1)_{\psi}$ and $U(1)_{\chi}$. In Eq.~(\ref{2}) the $U(1)_{N}$ gauge symmetry is defined as
\begin{equation}
U(1)_N=\dfrac{1}{4} U(1)_{\chi}+\dfrac{\sqrt{15}}{4} U(1)_{\psi}\,.
\label{3}
\end{equation}
In these $E_6$ inspired $U(1)_{N}$ extensions of the MSSM the right--handed neutrinos do not participate in the
gauge interactions and may be superheavy \cite{King:2005jy, King:2005my}. Therefore a see-saw mechanism can be used
to shed light on the origin of the mass hierarchy in the lepton sector, providing a comprehensive understanding
of the neutrino oscillations data. The successful leptogenesis is a distinctive feature of the $U(1)_N$ extensions
of the MSSM because the heavy Majorana right--handed neutrinos may decay into final states with lepton number $L=\pm 1$,
creating a lepton asymmetry in the early Universe~\cite{Hambye:2000bn,King:2008qb,Nevzorov:2017gir}.

In the $E_6$ inspired $U(1)_{N}$ extensions of the MSSM gauge anomalies get canceled if the particle spectrum
at low--energies contains complete representations of $E_6$. To ensure that the E$_6$SSM
is anomaly--free one is forced to extend the minimal matter content by extra matter beyond the MSSM
which, together with ordinary SM fermions, form three complete $27$--plets of $E_6$ ($27_i$ with $i=1,2,3$).
Each 27-dimensional representations of $E_6$ involves one generation of ordinary matter, a SM singlet field $S_i$,
that carries non--zero $U(1)_{N}$ charge, up- and down-type Higgs doublets $H^{u}_{i}$ and $H^{d}_{i}$ as well as
charged $\pm 1/3$ exotic quarks $D_i$ and $\bar{D}_i$. The presence of extra exotic matter may give
rise to non-diagonal flavor transitions and rapid proton decay. In the E$_6$SSM a set of discrete symmetries can
be used to suppress the corresponding operators \cite{King:2005jy,King:2005my}.

In this article we examine the dark matter-nucleon scattering cross section within the modification of the E$_6$SSM (SE$_6$SSM)
\cite{Nevzorov:2012hs,Athron:2014pua,Athron:2015vxg,Athron:2016gor} in which a single discrete $\tilde{Z}^{H}_2$ symmetry
forbids tree-level flavor-changing transitions, as well as the most dangerous baryon and lepton number violating operators.
The conservation of $\tilde{Z}^{H}_2$ symmetry and matter parity/$R$--parity implies the existence of at least two stable states
that may account for all or some of the observed cold dark matter density. At the same time null results from direct detection
experiments \cite{XENON:2018voc,PandaX-4T:2021bab,LUX-ZEPLIN:2022qhg} placed stringent limits on the dark matter-nucleon scattering cross section.
Here we explore a new variant of the SE$_6$SSM in which two stable states are gravitino and the lightest neutral exotic fermion.
In general, the spin-independent part of the dark matter-nucleon scattering cross section can be much larger in this case than the
corresponding experimental limit. Nevertheless the results of our analysis indicate that there is a part of the SE$_6$SSM parameter space,
in which this cross section is sufficiently strongly suppressed.

The paper is organised as follows. In the next section we specify a new variant of the SE$_6$SSM which is considered in this article.
In section 3 we investigate the dependence of the dark matter-nucleon scattering cross section on the parameters of the SE$_6$SSM.
Section 4 concludes the paper.

\section{The SE$_6$SSM}

Since 2006 several modifications of the E$_6$SSM have been explored
\cite{King:2005jy,King:2005my,Nevzorov:2012hs,Athron:2014pua,Howl:2007hq,Howl:2007zi,Howl:2008xz,Howl:2009ds,
Athron:2010zz,Hall:2011zq,Callaghan:2012rv,Callaghan:2013kaa,King:2016wep,Khalil:2020syr}.
The implications of the $U(1)_{N}$ extensions of the MSSM were considered for $Z$--$Z'$ mixing~\cite{Suematsu:1997au},
neutralino sector~\cite{Suematsu:1997au,Keith:1997zb,Keith:1996fv},
electroweak (EW) symmetry breaking (EWSB) \cite{Keith:1997zb,Suematsu:1994qm,Daikoku:2000ep},
the renormalization group (RG) flow of couplings~\cite{Keith:1997zb,King:2007uj},
the renormalization of vacuum expectation values (VEVs)~\cite{Sperling:2013eva,Sperling:2013xqa},
non-standard neutrino models~\cite{Ma:1995xk} and
dark matter~\cite{Khalil:2020syr,Hall:2009aj}.
Within the E$_6$SSM the upper bound on the lightest Higgs mass near the quasi-fixed point was examined
in~\cite{Nevzorov:2013ixa,Nevzorov:2015iya}. The corresponding quasi-fixed point is an intersection of
the invariant and quasi-fixed lines~\cite{Nevzorov:2001vj,Nevzorov:2002ub}.
The particle spectrum in the constrained E$_6$SSM (cE$_6$SSM) and its modifications was analyzed
in~\cite{Athron:2015vxg,Athron:2016gor,Athron:2011wu,Athron:2008np,Athron:2009ue,Athron:2009bs,Athron:2012sq}.
The degree of fine tuning and threshold corrections were explored in Refs.~\cite{Athron:2013ipa,Athron:2015tsa}
and \cite{Athron:2012pw} respectively.
Extra exotic matter in the E$_6$SSM may result in distinctive LHC signatures
\cite{King:2005jy,King:2005my,Howl:2007zi,Athron:2010zz,Athron:2011wu,
King:2006vu,King:2006rh,Belyaev:2012si,Belyaev:2012jz}
and can give rise to non-standard Higgs decays
\cite{Athron:2014pua,Nevzorov:2015iya,Nevzorov:2013tta,Hall:2010ix,Hall:2010ny,Hall:2011au,Hall:2013bua,Nevzorov:2014sha,Athron:2016usd}.

The SE$_6$SSM implies that below the scale $M_X$ three complete $27$--plets are
accompanied by SM singlet superfield $\phi$, which does not participate in the $E_6$ interactions,
and a set of pairs of supermultiplets $M_l$ and $\overline{M}_l$, which belong to
additional $27^{\prime}_l$ and $\overline{27}^{\prime}_l$ representations respectively.
Because $M_l$ and $\overline{M}_l$ carry opposite $SU(3)_C \times SU(2)_W \times U(1)_Y \times U(1)_N$
quantum numbers gauge anomalies still cancel. In the simplest case the set of $M_l$ and $\overline{M}_l$ involves
three pairs of $SU(2)_W$ doublets, i.e. $L_4$ and $\overline{L}_4$, $H_u$ and $\overline{H}_u$, $H_d$ and $\overline{H}_d$,
as well as a pair of superfields $S$ and $\overline{S}$. The field content of the SE$_6$SSM may originate from
the $E_6$ orbifold GUT model in six dimensions in which the appropriate splitting of the bulk
$27^{\prime}$ supermultiplets can be achieved \cite{Nevzorov:2012hs}.

The supermultiplets $\phi$, $S$, $\overline{S}$, $H_u$, $H_d$, $L_4$ and $\overline{L}_4$ are required to be even under
the $\tilde{Z}^{H}_2$ custodial symmetry whereas all other supermultiplets are odd \cite{Athron:2014pua}.
In the SE$_6$SSM superpotential the $\tilde{Z}^{H}_2$ symmetry allows the interactions that comes from $27'_l \times 27'_m \times 27'_n$
and $27'_l \times 27_i \times 27_k$ but it forbids all terms which originate from $27_i \times 27_j \times 27_k$,
where indexes $l,m,n$ are associated with the supermultiplets $M_l$ while family indexes $i,j,k=1,2,3$ run over three generations.
Such structure of the superpotential ensures that it does not involve any operators that give rise to rapid proton decay.
Because the set of supermultiplets $M_{l}$ includes only two Higgs doublets $H_d$ and $H_u$, the up-type quarks couple to
just $H_u$ whereas the down-type quarks and charged leptons couple to $H_d$ only. Therefore the flavor-changing processes are
suppressed in the SE$_6$SSM at tree-level.

In the simplest scenario $\overline{H}_u$ and $\overline{H}_d$ get combined with the superposition of the corresponding
components from the $27_i$, forming vectorlike states with masses of order $M_X$. The components of the supermultiplets $L_4$
and $\overline{L }_4$, as well as $S$ and $\overline{S}$ are expected to gain the TeV scale masses. The presence of $L_4$ and
$\overline{L}_4$ at low energies permits the lightest exotic quarks to decay within a reasonable time as well as
facilitates the gauge coupling unification \cite{King:2007uj} and the generation of the baryon asymmetry of
the Universe \cite{Nevzorov:2017gir}.

Using the method discussed in~\cite{Hesselbach:2007te,Hesselbach:2007ta,Hesselbach:2008vt}, it was revealed that
in the E$_6$SSM and its simplest modifications the lightest $R$-parity odd states (lightest SUSY particles) have to be lighter than $60-65~\mbox{GeV}$~\cite{Hall:2010ix,Hall:2010ny,Hall:2011au,Hall:2013bua}. These states are predominantly linear
superpositions of the fermion components of the superfields $S_{i}$. Although the couplings of these lightest exotic
fermions to the SM particles are very small the lightest SUSY particle (LSP) could account for some of the cold dark matter
relic density if the LSP had a mass close to half the $Z$ boson mass $M_Z/2$ \cite{Hall:2010ix}. However in this case the SM-like
Higgs boson decays mostly into the lightest exotic fermions whereas its other branching ratios are suppressed. LHC experiments
have already excluded such scenario. When the lightest exotic fermions are substantially lighter than $M_Z$,
the LSP annihilation cross section tends to be too small resulting in too large cold dark matter density.
In the simplest phenomenologically viable scenario the sparticle spectrum of the SE$_6$SSM includes the lightest SUSY particles
with masses which are much smaller than $1\,\mbox{eV}$. These lightest exotic fermions form hot dark matter in our Universe.
Nevertheless they give a minor contribution to the total density of dark matter. The existence of neutral fermions with tiny
masses may lead to very interesting implications for the neutrino physics \cite{Frere:1996gb}.

The variant of the SE$_6$SSM explored here implies that in addition to all states mentioned above the low energy matter content of
the model involves at least three $E_6$ singlet superfields $\phi_i$ which are odd under the $\tilde{Z}^{H}_2$ symmetry.
This allows to avoid the appearance of the lightest exotic fermions with tiny masses in the particle spectrum.
It is expected that the components of the supermultiplets
\begin{equation}
\begin{array}{c}
(Q_i,\,u^c_i,\,d^c_i,\,L_i,\,e^c_i)
+(D_i,\,\bar{D}_i) + S_{i} + \phi_i + (H^u_{\alpha},\,H^d_{\alpha})\\[0mm]
+L_4+\overline{L}_4+S+\overline{S}+H_u+H_d+\phi\,,
\end{array}
\label{4}
\end{equation}
where $\alpha=1,2$ and $i=1,2,3$, have masses either of the order of $10\,\mbox{TeV}$ or considerably lower.
The right-handed neutrino superfields $N^c_i$ are assumed to be much heavier then $10\,\mbox{TeV}$.
The $U(1)_Y$ and $U(1)_{N}$ charges of all matter supermultiplets in the SE$_6$SSM are summarised in Table~\ref{tab-0}.

\begin{table}[ht]
\centering
\begin{tabular}{|c|c|c|c|c|c|c|c|c|c|c|c|c|}
\hline
& $Q_i$ & $u^c_i$ & $d^c_i$ & $L_i, L_4$ & $e^c_i$ & $S_i, S$ & $H^u_{\alpha}, H_u$ & $H^d_{\alpha}, H_d$ & $D_i$ & $\overline{D}_i$ &
$\overline{L}_4$ & $\overline{S}$\\
\hline
$\sqrt{\frac{5}{3}}Q^{Y}_i$ & $\frac{1}{6}$ & $-\frac{2}{3}$ & $\frac{1}{3}$ & $-\frac{1}{2}$ & $1$ & $0$ & $\frac{1}{2}$ & $-\frac{1}{2}$ & $-\frac{1}{3}$ & $\frac{1}{3}$ & $\frac{1}{2}$ & $0$\\
\hline
$\sqrt{{40}}Q^{N}_i$ & $1$ & $1$ & $2$ & $2$ & $1$ & $5$ & $-2$ & $-3$ & $-2$ & $-3$ & $-2$ & $-5$ \\
\hline
\end{tabular}
\caption{The $U(1)_Y$ and $U(1)_{N}$ charges of matter supermultiplets in the SE$_6$SSM. The superfields $N^c_i$, $\phi_i$ and $\phi$
have zero $U(1)_Y$ and $U(1)_{N}$ charges. }
\label{tab-0}
\end{table}

Integrating out the right-handed neutrino superfields $N^c_i$ and neglecting all suppressed non-renormalisable interactions,
the low-energy superpotential of the new variant of the SE$_6$SSM can be written as
\begin{equation}
\begin{array}{c}
W_{\rm SE_6SSM} = \lambda S (H_u H_d) - \sigma \phi S \overline{S} +
\dfrac{\kappa}{3}\phi^3+\dfrac{\mu}{2}\phi^2+\Lambda\phi
+ \mu_L L_4\overline{L}_4+ \tilde{\sigma} \phi L_4\overline{L}_4 + W_{IH}\\[2mm]
+ \kappa_{ij} S (D_{i} \overline{D}_{j}) + g^D_{ij} (Q_i L_4) \overline{D}_j+ h^E_{i\alpha} e^c_{i} (H^d_{\alpha} L_4)
+ g_{ij} \phi_i \overline{L}_4 L_j + W_{\rm MSSM}(\mu=0)\,.
\end{array}
\label{5}
\end{equation}
In Eq.~(\ref{5}) the part of the superpotential $W_{IH}$ describes the
interactions of $\phi_i$, $S_i$, $H^u_{\alpha}$ and $H^d_{\alpha}$ with the
$\tilde{Z}_2^H$ even supermultiplets $\phi$, $S$, $\overline{S}$, $H_u$ and $H_d$
\begin{equation}
\begin{array}{c}
W_{IH} = \tilde{M}_{ij} \phi_i \phi_j + \tilde{\kappa}_{ij} \phi \phi_i \phi_j
+ \tilde{\lambda}_{ij} \overline{S} \phi_i S_j  + \lambda_{\alpha\beta} S (H^d_{\alpha} H^u_{\beta})\\[2mm]
+ \tilde{f}_{i\alpha} S_{i} (H^d_{\alpha} H_u) + f_{i\alpha} S_{i} (H_d H^u_{\alpha})\,.
\end{array}
\label{6}
\end{equation}

The sector responsible for the breakdown of the gauge symmetry in the SE$_6$SSM is formed by
the scalar components of $\phi$, $S$, $\overline{S}$, $H_u$ and $H_d$. In the limit $\sigma\to 0$
the $U(1)_N$ $D$-term contribution to the effective scalar potential can force the minimum
of such potential to be along the $D$-flat direction \cite{Kolda:1995iw}. Indeed, if the coupling
$\sigma$ vanishes the part of the scalar potential that depends on $S$ and $\overline{S}$ takes the form
\begin{equation}
V_S(S,\,\overline{S}) = m^2_S |S|^2 + m^2_{\overline{S}} |\overline{S}|^2
+\dfrac{Q_S^2 g^{\prime \, 2}_1}{2}\left(|S|^2-|\overline{S}|^2\right)^2\,,
\label{7}
\end{equation}
where $Q_S$ is the $U(1)_N$ charge of $S$ and $\overline{S}$, $g^{\prime}_1$ is the $U(1)_N$ gauge
coupling whereas $m_S^2$ and $m^2_{\overline{S}}$ are the soft SUSY breaking mass parameters.
The last term in Eq.~(\ref{7}) is associated with the $U(1)_N$ $D$-term contribution. For
$\langle S \rangle = \langle \overline{S} \rangle$ the quartic term vanishes.
If in this case $(m^2_S + m^2_{\overline{S}})<0$ then the scalar potential (\ref{7})
has a run--away direction $\langle S \rangle = \langle \overline{S} \rangle \to\infty$.
The $F$-term contribution to the scalar potential (\ref{7}) induced by the small coupling $\sigma$
stabilizes the run-away direction so that the SM singlet superfields acquire VEVs which are much larger
than the sparticle mass scale $M_S$, i.e.
\begin{equation}
\langle \phi \rangle \sim \langle S \rangle \simeq \langle \overline{S} \rangle \sim \dfrac{M_S}{\sigma}\,,
\label{71}
\end{equation}
resulting in an extremely heavy $Z'$ boson. All extra exotic states can be also rather heavy in this limit.

\begin{table}[ht]
\centering
\begin{tabular}{|c|c|c|c|c|}
\hline
&$Q_i, u^c_i, d^c_i, L_i, e^c_i, N^c_i$ & $\overline{D}_i, D_i, H^d_{\alpha}, H^u_{\alpha}, S_{i}, \phi_i$
& $H_d, H_u, S, \overline{S}, \phi$     & $L_4, \overline{L}_4$\\
\hline
$\tilde{Z}^{H}_2$  & $-$              & $-$                   & $+$       & $+$     \\
\hline
$Z_{2}^{M}$        & $-$              & $+$                   & $+$       & $-$     \\
\hline
$Z_{2}^{E}$        & $+$              & $-$                   & $+$       & $-$     \\
\hline
\end{tabular}
\caption{Transformation properties of different supermultiplets under the discrete symmetries $\tilde{Z}^H_2$,
$Z_{2}^{M}$ and $Z_{2}^{E}$. The signs $+$ and $-$ correspond to the states which are even and odd under different $Z_2$ symmetries.}
\label{tab-1}
\end{table}


The conservation of $Z_{2}^{M}$ and $\tilde{Z}^{H}_2$ symmetries implies that $R$--parity and
$Z_{2}^{E}$ symmetry are also conserved where $\tilde{Z}^{H}_2 = Z_{2}^{M}\times Z_{2}^{E}$ \cite{Nevzorov:2012hs}.
The transformation properties of different supermultiplets under the $\tilde{Z}^{H}_2$, $Z_{2}^{M}$
and $Z_{2}^{E}$ symmetries are summarized in Table~\ref{tab-1}. Here we assume that gravitino
is the lightest $R$--parity odd state. Therefore it has to be stable and potentially contributes
to the density of dark matter. There is a large class of models in which gravitino can be much
lighter than the superpartners of other particles \cite{Giudice:1998bp,Dubovsky:1999xc,Gherghetta:2000qt,Gherghetta:2000kr}.
Recently the cosmological implications of the gravitino with mass $m_{3/2}\sim \mbox{KeV}$ were discussed
in Ref.~\cite{Gu:2020ozv}. Since gravitino is the lightest SUSY particle with $Z_{2}^{E}=+1$
the lightest exotic state with $Z_{2}^{E}=-1$ must be absolutely stable as well \cite{Nevzorov:2012hs}.

The scalar components of the supermultiplets $\phi_i$, $S_i$, $H^u_{\alpha}$ and $H^d_{\alpha}$ do not develop VEVs.
Their fermion components compose the exotic chargino and neutralino states.
In the limit, when all components of $\phi_i$ are considerably heavier than
the bosons and fermions from the supermultiplets $S_i$, $H^u_{\alpha}$ and $H^d_{\alpha}$,
the superfields $\phi_i$ can be integrated out so that the part of the SE$_6$SSM superpotential $W_{IH}$
reduces to
\begin{equation}
\begin{array}{c}
W_{IH} \to \widetilde{W}_{IH}\simeq -\widetilde{\mu}_{ij} S_i S_j + \lambda_{\alpha\beta} S (H^d_{\alpha} H^u_{\beta})
+ \tilde{f}_{i\alpha} S_{i} (H^d_{\alpha} H_u) + f_{i\alpha} S_{i} (H_d H^u_{\alpha})+...\,.
\end{array}
\label{8}
\end{equation}
Hereafter we use the field basis in which $\widetilde{\mu}_{ij}=\widetilde{\mu}_{i}\,\delta_{ij}$ and
$\lambda_{\alpha\beta}=\lambda_{\alpha\alpha}\,\delta_{\alpha\beta}$.

In this article we focus on the scenarios in which the lightest exotic state with $Z_{2}^{E}=-1$ is predominantly
formed by the fermion components of the supermultiplets $H^u_{1}$ and $H^d_{1}$ while all sparticles except gravitino
and all other exotic states have masses which are much larger than $1\,\mbox{TeV}$. If $H^u_{1}$ and $H^d_{1}$
mostly interact with $H_u$, $H_d$ and $S_1$, whereas all other couplings of $H^u_{1}$ and $H^d_{1}$ are negligibly small,
the part of the mass matrix, that determines the masses of the lightest exotic neutralino states, can be written
in the following form
\begin{equation}
M^{ab}=-
\left(
\begin{array}{ccc}
0                                           & \mu_{11}                            & \dfrac{\tilde{f}_{11}}{\sqrt{2}} v_2 \\[3mm]
\mu_{11}                                    & 0                                   & \dfrac{f_{11}}{\sqrt{2}} v_1 \\[3mm]
\dfrac{\tilde{f}_{11}}{\sqrt{2}} v_2        & \dfrac{f_{11}}{\sqrt{2}} v_1        & \widetilde{\mu}_1             \\
\end{array}
\right)\,,
\label{9}
\end{equation}
where $\mu_{11}\simeq \lambda_{11} \langle S \rangle$, $v_1$ and $v_2$ are the VEVs of the Higgs doublets $H_d$ and $H_u$,
i.e. $\langle H_d \rangle = v_1/\sqrt{2}$ and $\langle H_u \rangle = v_2/\sqrt{2}$. Instead of $v_1$ and $v_2$ it is more convenient
to use $\tan\beta=v_2/v_1$ and $v=\sqrt{v_1^2+v_2^2} \approx 246\,\mbox{GeV}$. Hereafter we neglect the contribution of loop corrections
to the mass matrix (\ref{9}). In this case the mass of the charged fermion components of
$H^u_{1}$ and $H^d_{1}$ is determined by $\mu_{11}$, i.e. $m_{\chi^{\pm}_1}=\mu_{11}$.

When $|\widetilde{\mu}_1|$ is much larger than $|\mu_{11}|$ and $v$, the perturbation theory method can be used to diagonalise
the mass matrix (\ref{9}) (see, for example, \cite{Kovalenko:1998dc,Nevzorov:2000uv,Nevzorov:2001um,Nevzorov:2004ge}).
This method yields
\begin{equation}
\begin{array}{c}
m_{\chi_1} \simeq m_{\chi^{\pm}_1} - \Delta_1\,,\qquad\quad m_{\chi_2} \simeq m_{\chi^{\pm}_1} + \Delta_2\,,\qquad\quad
m_{\chi_2} \simeq \widetilde{\mu}_1 + \Delta_1 + \Delta_2\,,\\[2mm]
\Delta_1 \simeq \dfrac{(\tilde{f}_{11} v\sin\beta + f_{11} v\cos\beta)^2}{4(\widetilde{\mu}_1-m_{\chi^{\pm}_1})}\,,\qquad\qquad
\Delta_2 \simeq \dfrac{(\tilde{f}_{11} v\sin\beta - f_{11} v\cos\beta)^2}{4(\widetilde{\mu}_1+m_{\chi^{\pm}_1})}\,.
\end{array}
\label{10}
\end{equation}
From Eq.~(\ref{10}) it follows that the masses of the lightest exotic chargino and neutralino states
in the leading approximation are set by $\mu_{11}$. If the mass difference $m_{\chi_2}-m_{\chi_1}$
is as small as $O(100\,\mbox{KeV})$, the inelastic scattering processes $\chi_1 N\to \chi_2 N$, where $N$
is a nucleon, may occur. In this article we restrict our consideration to the part of the SE$_6$SSM
parameter space where $m_{\chi_2}-m_{\chi_1}> 200\,\mbox{MeV}$.  Because of this the inelastic
scattering processes do not take place.  Moreover the lifetime of $\chi_2$ is much smaller than
$1\,\mbox{sec}$ that allows to preserve the success of the Big Bang Nucleosynthesis (BBN).

In the scenarios under consideration the contribution of the lightest neutral exotic fermion to the cold dark matter
relic density can be estimated using the approximate formula
\begin{equation}
\Omega_{\tilde{H}} h^2 \simeq 0.1 \, \left(\dfrac{\mu_{11}}{1\,\mbox{TeV}}\right)^2\,.
\label{11}
\end{equation}
which was derived in the case of the Higgsino dark matter within the MSSM (see, for example, \cite{Arkani-Hamed:2006wnf,Chalons:2012xf}).
Because the Planck observations lead to \cite{Ade:2015xua}
\begin{equation}
(\Omega h^2)_{\text{exp.}} = 0.1188 \pm 0.0010\,,
\label{12}
\end{equation}
in the phenomenologically viable scenarios $\mu_{11}$ is expected to be lower than $1.1\,\mbox{TeV}$.
If $\mu_{11}\lesssim 1.1\,\mbox{TeV}$ then gravitino may account for some or major part of the observed cold
dark matter density.

In general, to find a viable cosmological scenario with stable gravitino one has to ensure that the decay
products of the lightest unstable $R$-parity odd (or exotic) particle $Y$ do not alter the abundances of light
elements induced by BBN. The decays of state $Y$ change the abundances of light elements
the more the longer its lifetime $\tau_Y$ is. This problem can be evaded if sparticle $Y$ decays before BBN, i.e.
$\tau_Y\lesssim 1\,\mbox{sec}$. The lifetime of state $Y$ decaying into its SM partner (or lightest
exotic fermion) and gravitino can be estimated as \cite{Feng:2004mt}
\begin{equation}
\tau_Y \sim 48\pi \dfrac{m_{3/2}^2 M_P^2}{m_Y^5}\,,
\label{121}
\end{equation}
where $m_Y$ is its mass and $M_P=(8\pi G_N)^{-1/2} \simeq 2.4\cdot 10^{18}\,\mbox{GeV}$ is the reduced Planck mass.
In order to get $\tau_Y\lesssim 1\,\mbox{sec}$ for $m_Y\simeq 1\,\mbox{TeV}$ we should restrict our consideration to
$m_{3/2}\lesssim 1\,\mbox{GeV}$.

If gravitinos mostly originate from scattering processes of particles in the thermal bath then their abundance is
approximately proportional to the reheating temperature $T_R$ after inflation. In the leading approximation one
finds \cite{Bolz:2000fu}
\begin{equation}
\Omega_{3/2} h^2 \sim 0.27 \left(\dfrac{T_R}{10^8 GeV}\right) \left(\dfrac{1\,\mbox{GeV}}{m _{3/2}}\right) \left(\dfrac{M_{\tilde{g}}}{1\,\mbox{TeV}}\right)^2\,,
\label{122}
\end{equation}
where $M_{\tilde{g}}$ is a gluino mass. Taking into account that $\Omega_{3/2} h^2\le 0.12$, for $M_{\tilde{g}}\gtrsim 2-3\,\mbox{TeV}$
and $m _{3/2}\simeq 1\,\mbox{GeV}$ one obtains an upper bound on the reheating temperature, i.e.
$T_R\lesssim 10^{6-7}\,\mbox{GeV}$ \cite{Hook:2018sai}. However even for so low reheating temperatures the appropriate amount
of the baryon asymmetry can be induced within the SE$_6$SSM via the decays of the lightest right--handed neutrino/sneutrino
into exotic states \cite{Nevzorov:2017gir}.

\section{Dark matter-nucleon scattering cross section}

Since the couplings of gravitino to the SM particles are negligibly small, in the scenarios under consideration
the interactions of the cold dark matter with the baryons are determined by the couplings of the lightest neutral
exotic fermion $\chi_1$. The low energy $\chi_1$--quark effective Lagrangian is
\begin{equation}
\mathcal{L}_{\chi_1 q}=\sum_q \Bigl( a_q \bar{\chi}_1\chi_1 \bar{q}q + d_q \bar{\chi}_1 \gamma^{\mu}\gamma_5 \chi_1 \bar{q} \gamma_{\mu}\gamma_5 q\Bigr)\,.
\label{14}
\end{equation}
The first term in the brackets gives rise to a spin--independent interaction while the second one
is associated with a spin--dependent interaction.

For spin--independent interactions experiments quote the cross--section to scatter off a nucleon
which is given by \cite{Ellis:2008hf,Kalinowski:2008iq}
\begin{equation}
\sigma_{SI}=\dfrac{4 m^2_r}{\pi} \dfrac{(Z f^p + (A-Z)f^n)^2}{A^2}\,,\qquad\qquad
m_r=\dfrac{m_{\chi_1} m_N}{m_{\chi_1}+m_N}\,,
\label{15}
\end{equation}
where $A$ and $Z$ are the nucleon number and charge of the target nucleus, whereas $f^p$ and $f^n$ are
related to the quantities entering $\mathcal{L}_{\chi_1 q}$. To simplify our analysis we assume
that $f^p \approx f^n \approx f^N$ while
\begin{equation}
\dfrac{f^N}{m_N}=\sum_{q=u,d,s} \dfrac{a_q}{m_q} f^N_{Tq} + \dfrac{2}{27} \sum_{Q=c,b,t} \dfrac{a_Q}{m_Q} f^N_{TQ}\,,
\label{16}
\end{equation}
$$
m_N f^N_{Tq} = \langle N | m_{q}\bar{q}q |N \rangle\,, \qquad\qquad f^N_{TQ} = 1 - \sum_{q=u,d,s} f^N_{Tq}\,.
$$
From Eqs.~(\ref{15})--(\ref{16}) one can see that $\sigma_{SI}$ depends rather strongly on the hadronic matrix elements,
i.e. the coefficients $f^N_{Tq}$, which are related to the $\pi$--nucleon $\sigma$ term and the spin content of the nucleon.
Here we set $f^N_{Tu}\simeq 0.0153$, $f^N_{Td}\simeq 0.0191$ and $f^N_{Ts}\simeq 0.0447$. These values of the hadronic matrix
elements are the default values used in micrOMEGAs, as determined in Ref.~\cite{Belanger:2013oya} from lattice results
(see also Ref.~\cite{Alarcon:2011zs,Thomas:2012tg,Cheng:2012qr,Alarcon:2012nr}).

Because in the SE$_6$SSM the exotic neutralino states do not couple to quarks and squarks, the coefficients $a_q$
receives only contributions from the $t$--channel exchange of CP--even Higgs bosons. In the scenarios under consideration
the mass of the lightest Higgs particle $m_{h_1}$ is much smaller than the masses of all other Higgs bosons.
Therefore in our analysis we ignore all contributions induced by the heavy Higgs exchange. Moreover in this case the lightest
Higgs scalar manifests itself in the interactions with the SM fermions and gauge bosons as a SM--like Higgs so that
\begin{equation}
\dfrac{a_q}{m_q} \simeq \dfrac{a_Q}{m_Q} \simeq \dfrac{g_{h\chi\chi}}{v m^2_{h_1}}\,.
\label{17}
\end{equation}
When $\Delta_1 \ll \mu_{11}\ll \widetilde{\mu}_1$ the coupling $g_{h\chi\chi}$ of
the SM-like Higgs to the lightest exotic neutralino is given by
\begin{equation}
|g_{h\chi\chi}|\simeq \dfrac{\Delta_1}{v}\,.
\label{18}
\end{equation}

The dominant contribution to the coefficients $d_q$ in Eq.~(\ref{14}) comes from $t$--channel Z boson exchange.
In the field basis $(\tilde{H}^{d0}_1,\,\tilde{H}^{u0}_1,\,\tilde{S}_1)$ the two lightest exotic neutralinos are
made up of the following superposition of interaction states
\begin{equation}
\chi_{\alpha} = N_{\alpha}^1 \tilde{H}^{d0}_1 + N_{\alpha}^2 \tilde{H}^{u0}_1 + N_{\alpha}^3 \tilde{S}_1\,,
\label{19}
\end{equation}
where $\alpha=1,2$.
In Eq.~(\ref{19}) $N^a_i$ is the exotic neutralino mixing matrix defined by
\begin{equation}
N_i^a M^{ab} N_j^b = m_i \delta_{ij}, \qquad\mbox{ no sum on } i.
\label{20}
\end{equation}
where $M^{ab}$ is $3\times 3$ exotic neutralino mass matrix given by Eq.~(\ref{9}).
Using the above compositions of the lightest exotic neutralino states (\ref{19}) it is straightforward to derive
the couplings of these states to the $Z$-boson. The part of the Lagrangian that describes the interactions
of $Z$ with $\chi_1$ and $\chi_2$ can be written as
\begin{equation}
\mathcal{L}_{Z\chi\chi}=\sum_{\alpha,\beta}\dfrac{M_Z}{2 v}Z_{\mu}
\biggl(\chi^{T}_{\alpha}\gamma_{\mu}\gamma_{5}\chi_{\beta}\biggr) R_{Z\alpha\beta}\,,\qquad\qquad
R_{Z\alpha\beta}=N_{\alpha}^1 N_{\beta}^1 - N_{\alpha}^2 N_{\beta}^2\,.
\label{21}
\end{equation}
Then the coefficients $d_q$ and the corresponding cross section can be presented in
the following form:
\begin{equation}
\sigma^{p,n}_{SD}=\dfrac{12 m^2_r}{\pi}\Bigg(\sum_{q=u,d,s} d_q \Delta^{p,n}_q\Biggr)^2\,,\qquad\qquad
d_q=\dfrac{T_{3q}}{2v^2}R_{Z11}\,,
\label{22}
\end{equation}
where $T_{3q}$ is the third component of isospin. For fractions of the nucleon spin carried by a
given quark $q$ we use (see also \cite{Chalons:2012xf})
\begin{equation}
\Delta^{p}_u=\Delta^{n}_d=0.842\,,\qquad \Delta^{p}_d=\Delta^{n}_u=-0.427\,,\qquad \Delta^{p}_s=\Delta^{n}_s=-0.085\,.
\label{23}
\end{equation}
In this article we examine the dependence of $\sigma_{SD}$ on the parameters of the SE$_6$SSM where we define
\begin{equation}
\sigma_{SD}=\dfrac{1}{2}\Biggl(\sigma^{p}_{SD} + \sigma^{n}_{SD}\Biggr)\,.
\label{24}
\end{equation}

Using the compositions of $\chi_1$ and $\chi_2$ one can obtain the analytical expressions for
the couplings of these states to the lightest Higgs boson, i.e.
\begin{equation}
g_{h\alpha\beta}= -\dfrac{1}{\sqrt{2}}\Biggl( f_{11} N^3_{\alpha} N^2_{\beta} \cos\beta +
\tilde{f}_{11} N^3_{\alpha} N^1_{\beta} \sin\beta \Biggr)\,.
\label{25}
\end{equation}
The perturbation theory method allows to compute the entries of the exotic neutralino mixing matrix $N^a_i$.
Substituting $N^a_i$ into Eq.~(\ref{25}) one can reproduce the analytical formula (\ref{18}) for
$g_{h11}=g_{h\chi\chi}$ and find the approximate expression for $R_{Z11}$. If $\widetilde{\mu}_1\gg \mu_{11}>0$
and $\mu_{11}$ is larger than $\tilde{f}_{11} v \sin\beta$ and $f_{11} v \cos\beta$, we get
\begin{equation}
R_{Z11}\simeq \dfrac{v^2(f_{11}^2\cos^2\beta - \tilde{f}_{11}^2 \sin^2\beta)}{4 \mu_{11} (\widetilde{\mu}_1 - \mu_{11})}\,.
\label{26}
\end{equation}

Here we restrict our considerations to low values of $\tan\beta\simeq 2$. In this case the particle spectrum
of the SE$_6$SSM may include the SM--like Higgs state with mass around 125~GeV if $\lambda \gtrsim g^{\prime}_1$.
Indeed, in this SUSY model the upper bound on the lightest Higgs boson mass $m_{h_1}$ is given by
\begin{equation}
\begin{array}{c}
m_{h_1}^2 \lesssim
\dfrac{\lambda^2}{2}v^2\sin^22\beta+ M_Z^2\cos^22\beta + g^{'2}_1 v^2 (Q_{H_d}\cos^2\beta + Q_{H_u}\sin^2\beta)^2+\tilde{\Delta}\,,
\end{array}
\label{27}
\end{equation}
where $Q_{H_d}$ and $Q_{H_u}$ are the $U(1)_N$ charges of $H_d$ and $H_u$ whereas $\tilde{\Delta}$ is the
contribution of loop corrections. When $\lambda\gtrsim \sqrt{2} (M_Z/v)\simeq 0.52 \gtrsim g^{\prime}_1$ the sum of the
first two terms can be larger than $M_Z^2$ for moderate values of $\tan\beta$. As a consequence for $\tan\beta\simeq 2$
the $125\,\mbox{GeV}$ Higgs state can be obtained. In this part of the SE$_6$SSM parameter space all Higgs states except
the lightest Higgs boson tend to have masses beyond the multi-TeV range and cannot be observed at the LHC experiments
\cite{King:2005jy,King:2005my,King:2006vu,King:2006rh}.

Within the SE$_6$SSM the masses of new exotic states including the $Z'$ boson are set by the VEVs of $S$ and $\overline{S}$.
LHC constraints require the extra $U(1)_{N}$ gauge boson to be heavier than $4.5\,\mbox{TeV}$ \cite{CMS:2021ctt,ATLAS:2019erb}.
When $\langle S \rangle \simeq \langle \overline{S} \rangle$ for the mass of the extra vector boson $M_{Z'}$ one finds
\begin{equation}
M_{Z'}\approx 2 g^{\prime}_1 Q_S \,\langle S \rangle\,.
\label{28}
\end{equation}
In order to get $M_{Z'}\gtrsim 4.5\,\mbox{TeV}$ we choose $\langle S \rangle \simeq \langle \overline{S} \rangle \gtrsim 6\,\mbox{TeV}$.
The experimental constraints on $Z-Z'$ mixing are also satisfied in this case. To avoid the experimental lower limit on the mass of the
lightest exotic chargino and to ensure that the lightest exotic neutralino state gives rise to the phenomenologically acceptable dark matter
density the interval of variation of $\mu_{11}$ is chosen so that $\mu_{11}\gtrsim 200\,\mbox{GeV}$ and $\mu_{11}\lesssim 1\,\mbox{TeV}$.
This corresponds to relatively small values of $\lambda_{11}\lesssim 0.17$. To simplify our analysis we set $\widetilde{\mu}_1\simeq 2\,\mbox{TeV}$.
In addition, we also require the validity of perturbation theory up to the scale $M_X$ that, in particular, constrains the allowed range of
$f_{11}$ and $\tilde{f}_{11}$.

\begin{figure}[h!]
\centering
\includegraphics[width=7.7cm]{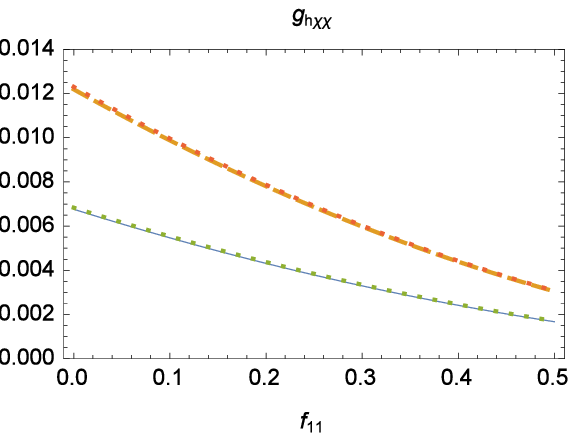}
\includegraphics[width=7.7cm]{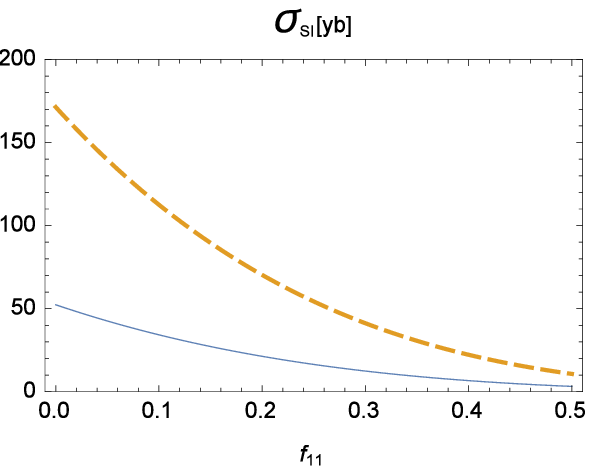}
\caption{({\bf Left}) The coupling $g_{h\chi\chi}$ and ({\bf Right}) the cross--section $\sigma_{SI}$
as a function of $f_{11}$ for $\tilde{f}_{11}=-0.5$, $\tan\beta=2$, $\widetilde{\mu}_1=2\,\mbox{TeV}$,
$\mu_{11} = 200\,\mbox{GeV}$ (solid lines) and $\mu_{11} = 1\,\mbox{TeV}$ (dashed lines). The dotted
lines correspond to the approximate expression for $g_{h\chi\chi}$ (\ref{18}).}
\label{fig1}
\end{figure}

The results of our analysis are summarised in Figs.~1-2. From the approximate expressions (\ref{18}) and (\ref{26}) it follows that
the couplings $g_{h\chi\chi}$ and $R_{Z11}$ vanish if $f_{11}\approx - \tilde{f}_{11} \tan\beta$. In other words the
interactions of the lightest exotic neutralino with the baryons can be extremely weak in some part of the SE$_6$SSM parameter space.
Our numerical analysis indicates that the perturbation theory method describes the dependence of $g_{h\chi\chi}$ and $R_{Z11}$
on the SE$_6$SSM parameter quite well when $\widetilde{\mu}_1$ is sufficiently large (see Figs.~1-2). In particular, $g_{h\chi\chi}$ and
$\sigma_{SI}$ grow whereas $R_{Z11}$ and $\sigma_{SD}$ diminish with increasing $\mu_{11}$ from $200\,\mbox{GeV}$ to $1\,\mbox{TeV}$.
Since in the scenario under consideration $\tilde{f}_{11}$ is chosen to be negative while $f_{11}$ is positive, $g_{h\chi\chi}$ and
$R_{Z11}$ as well as the spin--dependent and spin--independent cross sections decreases when $f_{11}$ grows.

\begin{figure}[h!]
\centering
\includegraphics[width=7.7cm]{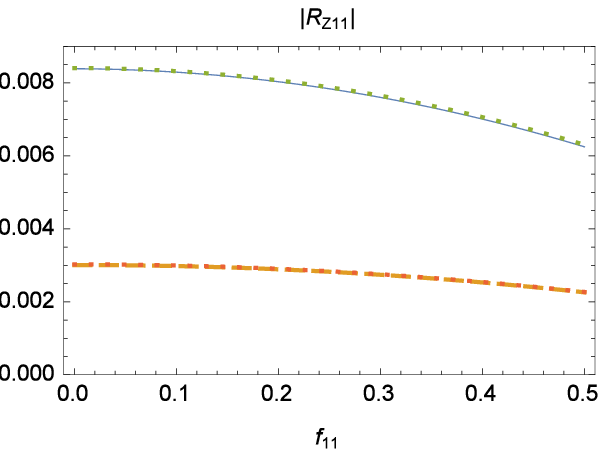}
\includegraphics[width=7.7cm]{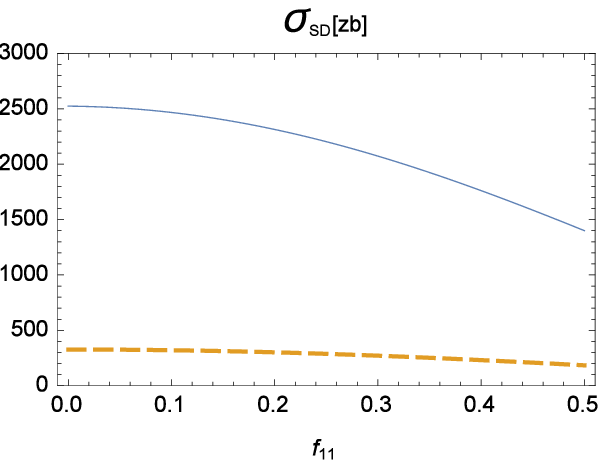}
\caption{({\bf Left}) The coupling $R_{Z11}$ and ({\bf Right}) the cross--section $\sigma_{SD}$
as a function of $f_{11}$ for $\tilde{f}_{11}=-0.5$, $\tan\beta=2$, $\widetilde{\mu}_1=2\,\mbox{TeV}$,
$\mu_{11} = 200\,\mbox{GeV}$ (solid lines) and $\mu_{11} = 1\,\mbox{TeV}$ (dashed lines). The dotted
lines correspond to the approximate expression for $R_{Z11}$ (\ref{26}).
}
\label{fig2}
\end{figure}

The spin--independent cross sections shown in Fig.~1 remain always smaller than $60\,\mbox{yb}$
for $\mu_{11}=200\,\mbox{GeV}$ and $300\,\mbox{yb}$ for $\mu_{11}=1\,\mbox{TeV}$ which are the
present upper bounds on $\sigma_{SI}$ set by the LZ experiment \cite{LUX-ZEPLIN:2022qhg}. In the part of the
SE$_6$SSM parameter space where $|g_{h\chi\chi}|$ and $|R_{Z11}|$ become smaller than $10^{-3}$ one cannot ignore
quantum corrections to the effective Lagrangian (\ref{14}) which are induced by the one--loop diagrams involving
the electroweak gauge bosons \cite{Hisano:2011cs,Hisano:2012wm}. The inclusion of such corrections results in
$\sigma_{SI}\sim 0.1\,\mbox{yb}$ even when $|g_{h\chi\chi}|\ll 10^{-3}$ at the tree level \cite{Nagata:2014wma}.

The values of $\sigma_{SD}$ and $\sigma_{SI}$ presented in Figs.~1 and 2 are substantially smaller than the maximal
possible values of the spin--dependent and spin--independent $\chi_1$--nucleon scattering cross sections in the SE$_6$SSM.
In the part of the parameter space examined in Figs.~1 and 2 the suppression of $\sigma_{SD}$ and $\sigma_{SI}$ is
caused by the large value of $\widetilde{\mu}_1$, which is associated with the sparticle mass scale, as well as
by the partial cancellations of different contributions to $g_{h\chi\chi}$ and $R_{Z11}$. The maximal possible values
of $\sigma_{SD}$ and $\sigma_{SI}$ are attained when $\mu_{11}\simeq \widetilde{\mu}_1$ and $\tilde{f}_{11} \sim f_{11}\sim 1$.
In this case the spin--independent $\chi_1$--nucleon scattering cross section can reach $20-30\,\mbox{zb}$
which is two orders of magnitude larger than the present experimental limit on $\sigma_{SI}$ \cite{LUX-ZEPLIN:2022qhg}.
However even when $\mu_{11}\simeq \widetilde{\mu}_1$, the desirable suppression of $\sigma_{SI}$ can be achieved if
$\tilde{f}_{11} \sim f_{11}\ll 0.1$.

\section{Conclusions}

To ensure anomaly cancellation the $U(1)_N$ extensions of the MSSM (E$_6$SSM) are expected to include at least
three $27$ representations of $E_6$ below the GUT scale $M_X$. These fundamental representations of $E_6$, in particular,
contain three families of Higgs--like doublets $H^{d}_{i}$ and $H^{u}_{i}$. One pair of such Higgs--like doublet supermultiplets
($H_u\equiv H^{u}_3$ and $H_d\equiv H^{d}_{3}$) gains VEVs resulting in the breakdown of the EW symmetry while
two other families ($H^{d}_{\alpha}$ and $H^{u}_{\alpha}$, where $\alpha=1,2$) are normally assumed to be inert.

In addition to three $27$--plets the low energy matter content of the SE$_6$SSM also involves an $E_6$ singlet superfield
$\phi$, a pair of superfields $S$ and $\overline{S}$ with opposite $U(1)_N$ charges as well as a pair of $SU(2)_W$
doublets $L_4$ and $\overline{L}_4$ with opposite $SU(2)_W \times U(1)_Y \times U(1)_N$ quantum numbers.
The supermultiplets $S$, $\overline{S}$, $L_4$ and $\overline{L}_4$ can come from additional $27^{\prime}_l$ and
$\overline{27}^{\prime}_l$. Within the SE$_6$SSM a single $\tilde{Z}^{H}_2$ symmetry prevents rapid proton decay
and tree-level non-diagonal flavor transitions. In this paper we have considered new variant of this SUSY model
in which the low energy particle spectrum is supplemented by three $E_6$ singlet superfields $\phi_i$ which are odd
under the $\tilde{Z}^{H}_2$ symmetry. This permits to avoid the presence of neutral exotic fermions with tiny masses
($\lesssim 1\,\mbox{eV}$) in the spectrum of the SE$_6$SSM. The scalar components of the $\tilde{Z}_2^H$ even superfields
$\phi$, $S$ and $\overline{S}$ may acquire very large VEVs, i.e.
$\langle\phi\rangle \sim \langle S\rangle \sim \langle \bar{S}\rangle \gg 10\,\mbox{TeV}$, breaking
the $U(1)_N$ gauge symmetry and generating masses of all extra exotic fermions which can be much heavier than all SM particles.

The conservation of $R$--parity and the $\tilde{Z}^{H}_2$ symmetry within the SE$_6$SSM ensures that at least
two neutral states can be stable giving rise to dark matter density. Since no firm indication of the
presence of dark matter has been observed at the direct detection experiments we assume that one of the stable
states is gravitino with mass $m_{3/2}\lesssim 1\,\mbox{GeV}$. Another stable state can be the lightest exotic fermion
which is mostly composed of the neutral fermion components of the supermultiplets $H^{d}_{1}$ and $H^{u}_{1}$.
In this case the lightest and second lightest exotic neutralino states ($\chi_1$ and $\chi_2$) as well as the lightest
exotic chargino $\chi^{\pm}_1$ are almost degenerate. If these exotic fermions have masses below $1.1\,\mbox{TeV}$ they
can lead to the phenomenologically acceptable density of the dark matter.

The interactions of dark matter with the baryons are defined by the couplings of $\chi_1$ because the interaction
of gravitino with the SM fermions and bosons is extremely weak. In this article we have explored the scenario in which
all scalars and exotic fermions except the lightest Higgs boson, gravitino, $\chi_1$, $\chi_2$ and $\chi^{\pm}_1$
have masses of the order of a few TeV or even higher. Therefore the spin--dependent and spin--independent
$\chi_1$--nucleon cross sections ($\sigma_{SD}$ and $\sigma_{SI}$) are dominated by the $t$-channel exchanges
of the $Z$ boson and the lightest Higgs scalar respectively. Although the maximal value of $\sigma_{SI}$ is about
$20-30\,\mbox{zb}$ in this SUSY model there is a part of the SE$_6$SSM parameter space in which
$\sigma_{SD}$ and $\sigma_{SI}$ can be strongly suppressed. In particular, the spin--independent dark matter--nucleon
scattering cross section can be considerably smaller than the corresponding experimental limit.

As mentioned before, the scenario under consideration implies that two lightest exotic neutralino states and the lightest
exotic chargino must be lighter than $1.1\,\mbox{TeV}$. Several collider experiments have searched for such
particles. However if the mass splitting between these states is small the decay products of $\chi_2$ and $\chi^{\pm}_1$
are very soft and may escape detection. This happens, for example, with the lightest ordinary neutralino and chargino
states within natural SUSY, where the splitting between the heavier neutralino or the chargino and the LSP is at least
a few GeV \cite{Baer:2012up,Baer:2012uy,Baer:2012cf}. In this case the results of the search for the lightest exotic
fermions depend on $\Delta = m_{\chi^{\pm}_1}-m_{\chi_1}$.

At the LHC the pair production of such states can occur through off-shell $W$ and $Z$ bosons.
ATLAS excluded the exotic chargino with masses below $193\,\mbox{GeV}$ and $140\,\mbox{GeV}$ for $\Delta \simeq 4.7\,\mbox{GeV}$
and $\Delta \simeq 2\,\mbox{GeV}$ respectively \cite{ATLAS:2019lng} while CMS ruled out such chargino with masses below $112\,\mbox{GeV}$
for $\Delta = 1\,\mbox{GeV}$ \cite{CMS:2019san}. On the other hand if $\Delta \lesssim 150\,\mbox{MeV}$ the exotic charginos may be long--lived.
When their lifetime is longer than the time needed to pass through the detector, these states appear as charged stable massive particles.
LHC experiments excluded such charginos with masses below $1090\,\mbox{GeV}$ \cite{ATLAS:2019gqq}.

The last experimental limit is not applicable in the case of the SE$_6$SSM scenario discussed here because $\Delta$ tends to be
larger than $300\,\mbox{MeV}$ \cite{Nagata:2014wma,Cirelli:2005uq}. The part of the SE$_6$SSM parameter space explored in Figs.~1 and 2
is associated with relatively small mass splitting between $\chi_1$, $\chi_2$ and $\chi^{\pm}_1$. For $m_{\chi^{\pm}_1}\simeq \mu_{11}= 200\,\mbox{GeV}$
the value of $\Delta\approx \Delta_1$ estimated using Eq.~(\ref{10}) changes from $1.67\,\mbox{GeV}$ to $0.42\,\mbox{GeV}$ when $f_{11}$ increases
from $0$ to $0.5$. As a consequence the lightest exotic chargino decays mainly into hadrons and the lightest exotic neutralino so that
it seems to be rather problematic to discover the set of these states at hadron colliders.
The discovery prospects for such fermions at future International Linear Collider look more promising (for a review see \cite{Baer:2013cma}).

Thus in this article we argued that within the new variant of the SE$_6$SSM, in which the cold dark matter density is formed by two stable
states, there are some regions of the parameter space that are safe from all current constraints. The SUSY model under consideration predicts
the existence of exotic fermions and exotic scalars which are colored and carry baryon and lepton numbers simultaneously \cite{Nevzorov:2012hs}.
In collider experiments such exotic quarks/squarks can only be created in pairs. The stringent LHC lower limits on the masses of scalar
leptoquarks \cite{ATLAS:2020dsk,ATLAS:2021oiz,CMS:2019ybf} are not directly applicable in the case of the SE$_6$SSM.
Indeed, in the $E_6$ inspired SUSY models ordinary scalar leptoquark can decay either to an up--type quark $u_i$ and charged lepton $\ell_j$
or to down--type quark $d_i$ and left--handed neutrino $\nu_j$. On the other hand in the SE$_6$SSM the lightest exotic colored particle $D_1$
(scalar or fermion) is odd under $\tilde{Z}^{E}_2$ symmetry. Therefore its decay always gives rise to the missing energy and transverse momentum
in the final state associated with the lightest exotic fermion, i.e.
$$
D_1 \to u_i (d_i) + \ell_j (\nu_j) + E^{\rm miss}_{T} + X\,.
$$
Here $X$ may include extra charged leptons and/or jets that can come from the decays of intermediate states. Since
at the LHC $D_1$ are pair produced the lightest exotic colored states may lead to some enhancement of the cross sections of
$pp\to jj \ell^{+}\ell^{-} + E^{\rm miss}_{T} + X$ and $pp\to jj + E^{\rm miss}_{T} + X$ if they are relatively light.
The discovery of these exotic colored states will provide a smoking gun signal of the model allowing to distinguish it from
other extensions of the SM.


\section*{Acknowledgements}
R.N. thanks X.~Tata for very valuable comments and remarks.
R.N. acknowledges fruitful discussions with S.~Khalil, S.~Moretti, M.~Vasiliev, M.~Vysotsky, H.~Waltari, E.~Zhemchugov and P.~V.~Zinin.

\end{document}